\documentclass[journal]{IEEEtran}

\IEEEoverridecommandlockouts
\usepackage{graphicx}
\usepackage{subfigure}
\usepackage{tikz}
\usepackage{pgfplots}
\usepackage{textcomp}
\usepackage{graphicx}
\usepackage{epstopdf}
\DeclareGraphicsExtensions{.png,.pdf}
\usepackage{amsfonts,latexsym,amssymb,shadow,amsmath,wrapfig,eepicemu}%,psfrag}
\usepackage[margin=0pt,font=small,labelfont=bf,justification=raggedright,format=hang]{caption}
\usepackage{xcolor}
\newcommand{\beq}{\begin{equation}}
\newcommand{\eeq}{\end{equation}}
\newcommand{\beqa}{\begin{eqnarray}}
\newcommand{\eeqa}{\end{eqnarray}}
\newcommand{\beqan}{\begin{eqnarray*}}
\newcommand{\eeqan}{\end{eqnarray*}}

\newcounter{l1}
\newcounter{l2}
\newcounter{l3}
\newcommand{\bdotlist}{\begin{list}{$\bullet$}{}}
\newcommand{\bboxlist}{\begin{list}{$\Box$}{}}
\newcommand{\bbboxlist}{\begin{list}{\raisebox{.005in}{{\tiny $\blacksquare$ \ \ }}}{}}
\newcommand{\bdashlist}{\begin{list}{$-$}{} }
\newcommand{\blist}{\begin{list}{}{} }
\newcommand{\barablist}{\begin{list}{\arabic{l1}}{\usecounter{l1}}}
\newcommand{\balphlist}{\begin{list}{(\alph{l2})}{\usecounter{l2}}}
\newcommand{\bAlphlist}{\begin{list}{\Alph{l2}.}{\usecounter{l2}}}
\newcommand{\bdiamlist}{\begin{list}{$\diamond$}{}}
\newcommand{\bromalist}{\begin{list}{(\roman{l3})}{\usecounter{l3}}}

\newtheorem{theorem}{Theorem}[section]
\newtheorem{exercise}[theorem]{Exercise}
\newtheorem{lemma}[theorem]{Lemma}
\newtheorem{proposition}[theorem]{Proposition}
\newtheorem{corollary}[theorem]{Corollary}
\newtheorem{definition}[theorem]{Definition}
\newtheorem{remark}[theorem]{Remark}
\newtheorem{example}[theorem]{Example}

\usepackage{cite}
\usepackage{amsmath,amssymb,amsfonts}
\usepackage{algorithmic,tcolorbox}
\usepackage{graphicx}
\usepackage{textcomp}
\usepackage{setspace}
\usepackage{url}
\usepackage{verbatim}
\usepackage{float}
\allowdisplaybreaks[4]
\usepackage{hyperref}
\usepackage{pgfplotstable}
\pgfplotsset{
    legend image code/.code={
        \draw [#1] (0cm,-0.1cm) rectangle (0.6cm,0.1cm);
    },
}
\def\BibTeX{{\rm B\kern-.05em{\sc i\kern-.025em b}\kern-.08em
    T\kern-.1667em\lower.7ex\hbox{E}\kern-.125emX}}
\pgfplotsset{compat=1.14}
\newtheorem{fact}{Fact}

\begin{document}

\title{Optimal Electricity Storage Sharing Mechanism\\ for Single Peaked Time-of-Use Pricing Scheme}
\author{Kui Wang, Yang Yu, \emph{and} Chenye Wu %\vspace{-0.6cm}
\thanks{The authors are with the Institute for Interdisciplinary Information Sciences (IIIS), Tsinghua University, Beijing, China, 100084. C. Wu is the correspondence author. Email: chenyewu@tsinghua.edu.cn.}}

\maketitle

\begin{abstract}
    Sharing economy has disrupted many industries. We foresee that electricity storage systems could be the enabler for sharing economy in electricity sector, though its implementation is a delicate task. Unlike in the 2-tier Time-of-Use (ToU) pricing, where greedy arbitrage policy can achieve the maximal electricity bill savings, most existing ToU schemes consist of multiple tiers, which renders the arbitrage challenging. The difficulty comes from the hedging against multiple tiers and the coupling between the decisions across the day. In this work, we focus on designing the energy sharing mechanism for single peaked ToU scheme. To solve the problem, we identify  that it suffices to understand the arbitrage policies for two forms of 3-tier ToU schemes. We submit that under mild conditions, the sharing mechanism yields a unique equilibrium, which supports the maximal social welfare.
\end{abstract}


\section{Introduction}
Sharing economy exploits huge Internet's value to heavy financial-cost yet idle assets \cite{horton2016owning}. This emerging business model is already disruptive for large industries such as transportation, accommodation and micro finance\cite{PwCsharinganalysis}. We notice that there are tremendous idle assets on the power grid, which already make the sharing economy business model attractive in electricity economics. However, the operational complexity due to physical constraints leads to computational challenges in the sharing market design and operation for the idle assets.

The current exploration on the sharing opportunities in electricity sector concentrates on real and virtual demand side assets. All relative studies demonstrate that such sharing opportunities need to be supported by three pillars: appropriate control policy with physical constraints, smart market design inducing incentive compatibility, and efficient algorithms to find the Pareto equilibrium.

%Net metering for rooftop photovoltaic (PV) panels has been thought of as the first sharing economy in electricity sector. However, it hasn't fully exploited the advantage of market due to its fixed price set by utilities. Also, net metering hasn't allowed technology to make sharing assets cheaper and easier, which is crucial for the new business model to be adopted on a much larger scale.

We imagine the first adopters could be to share the electricity storage in behind-the-meter setting \cite{8025410}. This is because the major concern in electricity sector comes from the regulatory uncertainty, and behind-the-meter setting is outside the purview of the utility. In such a setting, the firms could utilize their storage systems to arbitrage against the ToU pricing scheme, and share the excess energy via a local spot market. {In essence, this opportunity requires each firm to better utilize the unused capacity in its storage system, which motivates us to design the electricity storage sharing mechanism.}

It still warrants significant efforts to bring sharing electricity storage to practice. Wu \emph{et al.} propose a stylized model for storage sharing where in an industrial park, each firm faces 2-tier ToU pricing scheme \cite{wu2016sharing}. This stylized model elegantly lays out the theoretical foundation for the sharing economy business model. We further this result by generalizing from 2-tier to practical multi-tier ToU schemes. It is challenging even for a single firm's decision making, since {it needs to hedge} against multiple tiers and its decisions across the day are closely coupled together. The sharing market will emerge not only during peak but also partial peak periods. Hence, the market structure is more complex for examining the equilibrium behavior. In this work, we try to understand the system behavior (including individual's decision making as well as the {Nash equilibrium (N.E.)} behavior in various sharing markets) in single peaked\footnote{Single peaked ToU scheme: Each day, the rate first increases from off peak to peak, and then decreases to next off peak period.} ToU scheme. {We offer explicit formula to characterize the spot prices in the sharing markets.}

\vspace{-0.3cm}
\subsection{Related Work}
\vspace{-0.2cm}
Researchers have designed control policies for different pricing schemes. Wu and Yu {proposed} an optimal policy to arbitrage against 3-tier ToU pricing {to maximize the arbitrage profits in} \cite{3tiercontrol}. For dynamic pricing, Qin \emph{et al.} {proposed} an online modified greedy algorithm and {proved} its sub-optimality compared to offline \cite{qin2016online}. Van de ven \emph{et al.} {proposed} an optimal control policy for storage charging and discharging under Markovian random rates and demands \cite{6477197}. To the best of our knowledge, optimal control policy for general ToU has not been fully investigated.

The literature on sharing economy in electricity sector emerges only recently, most of which {investigated} the cooperation of consumers to hedge against stochastic risks. For example, Zhao \emph{et al.} {introduced} the optimal risky power contracts to the aggregation of multiple wind power plants for better bidding against 2-settlement markets in \cite{6913576}. Bitar \emph{et al.} {explored} the quantity risk sharing opportunities among wind power plants in nodal pricing scheme \cite{bitar2012optimal}. Chakraborty \emph{et al.} {introduced} the rooftop PV sharing mechanism for active users in \cite{chakraborty2018analysis}. Perera \emph{et al.} {proposed} a solar power sharing mechanism while avoiding the high voltage impact to grid \cite{perera2013power}. Zhao and Khazaei {designed} a cooperative game to aggregate multiple renewable power plants and {proved} that the core supports the social welfare \cite{7741403}. %The cooperation of these power plants can also be formed by a competitive game \cite{khazaei2019indirect}. 

As for storage sharing, Tushar \emph{et al.} {proposed} an auction-based storage sharing mechanism, where a unique equilibrium exists \cite{7387779}. Chakraborty \emph{et al.} {cast} the storage sharing problem in 2-tier ToU pricing as a coalitional game \cite{chakraborty2018sharing}. Different from previous works, we seek to design {spot markets enabling the storage sharing, and offer the explicit characterization for the sharing prices.}
%The work most related to ours is \cite{8025410}, where Kalathil \emph{et al.} analyze electricity storage sharing under 2-tier ToU pricing. We further the model to enable storage sharing under single peaked ToU.
\vspace{-0.3cm}
\subsection{Our Contributions}
\vspace{-0.2cm}
{
In this work, we significantly extend our previous works \cite{wu2016sharing} and \cite{3tiercontrol} by designing the optimal storage sharing mechanism for single peaked multi-tier ToU scheme {to minimize all firms' total cost for energy utilization}. In the sharing mechanism, we analytically examine the sharing prices in spot markets, and each individual's optimal decision making {to minimize its own cost}. In summary, our principal contribution can be summarized as follows:
}
\begin{itemize}
    \item \emph{Optimal control policy:} We propose the optimal control policy for standalone firm to arbitrage against single peaked ToU pricing scheme. Based on this control policy, we solve the individual optimal investment decision-making problem.
    \item \emph{Sharing spot market:} We identify that sharing could happen in all non-off peak periods. For each period, {we identify the aggregator-firms interaction game, and characterize its N.E.}. {The {N.E.} consists of the sharing prices and each individual's decision making. We also provide} the sufficient conditions for the existence of N.E.. 
    \item \emph{Performance assessment:} We prove that the outcome of our proposed sharing mechanism supports the maximal social welfare. We use simulation to compare our sharing mechanism and alternatives.
\end{itemize}

%\red{Reviewer says the contribution should be re cleared.}
The rest of the paper is organized as follows. Section \ref{ses} introduces the system model. Then, we propose the individual optimal control policy for arbitrage in Section \ref{Control}. Based on this control policy, we identify the market structure in Section \ref{sharingmechanism}. Then, Section \ref{3T} derives the optimal sharing mechanism for three-tier ToU scheme. Inspired by the intuition from Section \ref{3T}, we fully solve the problem for general single peaked ToU scheme in Section \ref{General}. Simulation verifies the performance of our proposed mechanism in Section \ref{Simulation}. Section \ref{Conclusionsection} gives concluding remarks and points out future work.

We provide all the necessary proofs in the Appendix.
%Due to space limits, all the proofs are provided in \cite{S_M}.

\section{Sharing Electricity Storage}
\label{ses}
\vspace{-0.2cm}
%\red{We cannot assume all the readers are experts. In the problem formulation, we should make our setup crystal clear. In the technical session (those theorems), we can omit many details and focus on the intuitions, by postponing the rigorous discussion in the Appendix. This should be the guideline to write any technical article.}

Assume a single peaked multi-tier ToU is implemented in an industrial park, where firms can invest storage for hedging against the price risks. To simplify the analysis, we suppose that an aggregator in the park coordinates the storage sharing and energy management, such as procuring energy from the grid. %Figure \ref{fig:parks} show 2 possible structures of the industrial park.
%We find that the analysis for single peaked ToU pricing depends on two basic forms of three-tier ToU schemes as shown in Fig. \ref{fig:ToU_3}. Hence, in the subsequent analysis, we will first focus on these two basic forms, and then synthesis the results for the general scheme. 

To obtain more insights through a stylized model, we make the following assumptions: 

\begin{enumerate}
    \item Arbitrage opportunity exists: the largest rate gap among hours is greater than the amortized cost of the storage investment.
    \item All players are price takers {of ToU prices}.
    \item All users' hourly energy consumption is inelastic.
    \item Demands in different periods are independent.
    %\item Statistical assumptions: the probability density function of demands in all non off peak periods of all firms is continuously differentiable, and total demand in each non off peak period is ranged from 0 to a sufficient large amount.
    \item Energy loss in storage operation is negligible. 
    \item Firms make investment decisions simultaneously.
\end{enumerate}
{
Assumption 1 is already practical for many ToU pricing schemes thanks to the decreasing cost for storage systems. For example, the amortized cost for Tesla Power 2.0 Lithium-ion over its 10-year lifetime is only 14 \textcent/kWh per day \cite{Tesla}. Assumption 2 implies that the user behaviors are not large enough to influence the ToU prices. This is reasonable for our industrial park setting. However, we want to emphasize that, when designing the sharing prices in our main results, we do consider all firms compete in the energy sharing market. Assumption 3 implies that without storage systems, all the firms have already fully exploited their own flexibilities in response to ToU prices. This allows us to focus on the additional benefits brought by the storage systems and establish a stylized model. 
We use numerical study to highlight that the performance of our scheme is close to that of the offline optimal with real data under assumption 4.
%This suggests assumption 4 does not make our analytical conclusions impractical.
The next assumption is temporary for more intuition during the analysis. We drop this assumption in Appendix A for general conclusions. 
The last one is the common technical assumption for game theoretic analysis. 
}

\section{Optimal Control for Standalone Firm}
\label{Control}
\vspace{-0.2cm}
The firms make sharing decisions according to their energy consumption patterns and storage control. Thus, the storage sharing market design and arrangement must be built upon the understanding of individual's decision. The individual control policy in 3-tier ToU scheme has been clarified by \cite{3tiercontrol}. {We first summarize this result and then} examine the individual storage control policy for a general single peaked ToU scheme.

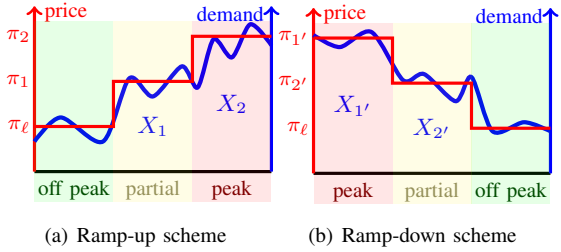
\begin{figure}[t]
\centering
\subfigure[Ramp-up scheme]{
\begin{tikzpicture}[xscale=0.7, yscale = 0.4]
%increasing mode
\draw[very thick] (0,0) -- coordinate (x axis mid) (4.5,0);
\node[align=right] at (2,-1) {};
\node[text=red] at (0.6,5.3) {\footnotesize price};
\node[text=blue] at (3.72,5.25) {\footnotesize demand};
\draw[very thick,red ,->] (0,0) -- coordinate (y axis mid) (0,5.5);
\draw[very thick,blue,->] (4.5,0) -- coordinate (y axis mid) (4.5,5.5);	
%\draw[blue, ultra thick, domain=0:1.3, samples=200, smooth, thick] plot (\x, {1.1+ 0.2*(cos(500*\x) + sin(100*\x^2) + exp(-0.5*\x) )});
%\draw[blue, ultra thick, domain=1.7:3, samples=150, smooth, thick] plot (\x, {2.7 - %0.2*(sin(400*\x) + sin(90*\x^2) - exp(-0.3*\x) )});
%\draw[blue, ultra thick, domain=3:4.5, samples=100, smooth, thick] plot (\x, {4.5 - 0.2*(sin(400*\x) + sin(90*\x^2) - exp(-0.3*\x) )});
%\addplot[color=blue, smooth] coordinates {
%    (0,1.5) (0.5,0.75) (1,1.8) (1.5,3.1) (2,3.3) (2.5,3.2) (3,2,7) (3.5,2,8) (4,4.6) (4.5,4.3)};
\draw [blue, ultra thick] plot [smooth] coordinates { (0,1) (0.5,1.8) (1.3, 1) (1.8,3.1) (2.25, 2.5) (2.8, 3.5) (3.1, 2.8) (3.4, 4.4) (3.75, 3.8) (4.1, 4.9) (4.5,4.2) };
%X's background
%\draw [very thick,gray, smooth] (1.5,1.5) --  (1.97,2.91);
%\draw [very thick,gray, smooth] (1.5,0.6) --  (2.14,2.52);
%\draw [very thick,gray, smooth] (1.6,0) --  (2.7,3.3);
%\draw [very thick,gray, smooth] (1.9,0) --  (2.95,3.15);
%\draw [very thick,gray, smooth] (2.2,0) --  (3,2.4);
%\draw [very thick,gray, smooth] (2.5,0) --  (3,1.5);
%\draw [very thick,gray, smooth] (2.8,0) --  (3,0.6);
\node[align=center, text=blue] at (2.25,1.5) {\small $X_1$};
%Y's background
%\draw [very thick,gray, dashed, smooth] (3,0.9) --  (3.3,0);
%\draw [very thick,gray, dashed, smooth] (3,1.8) --  (3.6,0);
%\draw [very thick,gray, dashed, smooth] (3,2.7) --  (3.9,0);
%\draw [very thick,gray, dashed, smooth] (3.2,3) --  (4.2,0);
%\draw [very thick,gray, dashed, smooth] (3.3,3.6) --  (3.65,2.55);
%\draw [very thick,gray, dashed, smooth] (3.85,1.95) --  (4.5,0);
%\draw [very thick,gray, dashed, smooth] (3.4,4.2) --  (4.5,0.9);
%\draw [very thick,gray, dashed, smooth] (3.8,3.9) --  (4.5,1.8);
%\draw [very thick,gray, dashed, smooth] (3.95,4.35) --  (4.5,2.7);
%\draw [very thick,gray, dashed, smooth] (4.1,4.8) --  (4.5,3.6);
\node[align=center, text=blue] at (3.75,2.25) {\small $X_2$};
%\draw[blue, ultra thick, domain=1.3:1.7, samples=100, smooth, thick] plot ([1.3, 1.7],[1, 2]);
%\node[align=center, text=blue] at (1,3.5) {Energy $X$};
%\node[align=center, text=blue] at (2.75,3.5) {Energy $Y$};
%\node[align=center, text=blue] at (4,3.5) {Energy $Z$};
\draw [draw=none, fill=green, fill opacity = 0.1] (0,-1) rectangle (1.5,5);
\draw [draw=none, fill=yellow, fill opacity = .1] (1.5,-1) rectangle (3,5);
\draw [draw=none, fill=red, fill opacity = .1] (3,-1) rectangle (4.5,5);
\draw [very thick,red, smooth] (0,1.5) --  (1.5,1.5)--(1.5,3)--(3,3)-- (3,4.5) -- (4.5,4.5);
%\draw [very thick,blue] (3.5,2.45) -- (3.5,2.25) ;
\node[anchor=north, text=black!70!green] at (0.75,0) {\footnotesize off peak};
\node[anchor=north, text=black!50!yellow] at (2.25,0) {\footnotesize partial};
\node[anchor=north, text=black!50!red] at (3.75,0) {\footnotesize peak};
%\node[anchor=north, text=black!50!yellow] at (4.5,-0.1) {partial};
\node[anchor=east,text=red] at (0.1,1.5) {\footnotesize $\pi_\ell$};
\node[anchor=east,text=red] at (0.1,3) {\footnotesize $\pi_1$};
\node[anchor=east,text=red] at (0.1,4.5) {\footnotesize $\pi_2$};
%\node[anchor=west] at (0.3,3.5) {Pr $\{S \geq s\} = \rho$};
\end{tikzpicture}
}
%\hspace{0in}
\hskip -12pt
\subfigure[Ramp-down scheme]{
\begin{tikzpicture}[xscale=0.7, yscale = 0.4]
%decreasing mode
\draw[very thick] (0,0) -- coordinate (x axis mid) (4.5,0);
%\node[align=right] at (2,-1) {};
\node[text=red] at (0.6,5.3) {\footnotesize price};
\node[text=blue] at (3.72,5.25) {\footnotesize demand};
\draw[very thick,red ,->] (0,0) -- coordinate (y axis mid) (0,5.5);
\draw[very thick,blue,->] (4.5,0) -- coordinate (y axis mid) (4.5,5.5);	
%\draw[blue, ultra thick, domain=0:2, samples=200, smooth, thick] plot (\x, {1.1+ 0.2*(cos(500*\x) + sin(100*\x^2) + exp(-0.5*\x) )});
%\draw[blue, ultra thick, domain=2:3.5, samples=150, smooth, thick] plot (\x, {2.3 - 0.2*(sin(400*\x) + sin(90*\x^2) - exp(-0.3*\x) )});
%\draw[blue, ultra thick, domain=3.5:4.5, samples=100, smooth, thick] plot (\x, {1.5 - 0.2*(sin(400*\x) + sin(90*\x^2) - exp(-0.3*\x) )});
\draw [blue, ultra thick] plot [smooth] coordinates { (0,4.6) (0.5,4.2) (1.1, 4.7) (1.7,3.1) (2.1, 3.3) (2.7, 2.4) (3, 3.2) (3.4, 1.4) (4, 1.7) (4.5,1.4) };
\node[align=center, text=blue] at (0.75,2.25) {\small $X_{1'}$};
\node[align=center, text=blue] at (2.25,1.5) {\small $X_{2'}$};
%\node[align=center, text=blue] at (1,3.5) {Energy $X$};
%\node[align=center, text=blue] at (2.75,3.5) {Energy $Y$};
%\node[align=center, text=blue] at (4,3.5) {Energy $Z$};
\draw [draw=none, fill=red, fill opacity = 0.1] (0,-1) rectangle (1.5,5);
\draw [draw=none, fill=yellow, fill opacity = .1] (1.5,-1) rectangle (3,5);
\draw [draw=none, fill=green, fill opacity = .1] (3,-1) rectangle (4.5,5);
\draw [very thick,red] (0,4.5) --  (1.5,4.5)--(1.5,3)--(3,3)-- (3,1.5) -- (4.5,1.5);
%\draw [very thick,blue] (3.5,2.45) -- (3.5,2.25) ;
\node[anchor=north, text=black!50!red] at (0.75,0) {\footnotesize peak};
\node[anchor=north, text=black!50!yellow] at (2.25,0) {\footnotesize partial};
\node[anchor=north, text=black!70!green] at (3.75,0) {\footnotesize off peak};
%\node[anchor=north, text=black!50!yellow] at (4.5,-0.1) {partial};
\node[anchor=east,text=red] at (0.1,1.5) {\footnotesize $\pi_\ell$};
\node[anchor=east,text=red] at (0.1,3) {\footnotesize $\pi_{2'}$};
\node[anchor=east,text=red] at (0.1,4.5) {\footnotesize $\pi_{1'}$};
%\node[anchor=west] at (0.3,3.5) {Pr $\{S \geq s\} = \rho$};
\end{tikzpicture}    
}
\caption{Two basic 3-tier ToU pricing.\vspace{-0.3cm}}
\label{fig:ToU_3}
\end{figure}

\vspace{-0.3cm}
\subsection{$(M,C)$ Storage Control Policy for 3-tier ToU Scheme}
\vspace{-0.2cm}
The analysis for 3-tier ToU scheme in \cite{3tiercontrol} reveals the nature of the {storage control} problems in the single peaked ToU schemes: there are two basic schemes of storage control shown in Fig. \ref{fig:ToU_3}. The optimal control for ramp-down scheme ({as shown in} Fig. \ref{fig:ToU_3} (b)) is a greedy policy: fully charge the battery during the off peak period and maximally use the stored energy sequentially from the peak to the partial peak period.

The optimal storage control for ramp-up scheme ({as shown in} Fig. \ref{fig:ToU_3} (a)) is refereed to as the  $(M,C)$ control policy in \cite{3tiercontrol}, where $C$ denotes the storage capacity and $M$ denotes the energy reserved for peak demand use.

Denote the random partial peak and peak demand by $X_1$ and $X_2$, respectively. $F_{X_2}(\cdot)$ is the cumulative density function (\emph{cdf}) of random variable $X_2$; $F_{X_1+X_2|X_2>M^*}$ is the conditional \emph{cdf} of $X_1+X_2$ given $X_2 > M^*$; $\pi_h$, $\pi_m$, {$\pi_\ell$} are the peak, partial peak and off peak rates, respectively; $\pi_s$ is the amortized investment cost of storage. Then, the optimal parameters for $(M,C)$ control policy can be determined as follows:
\begin{equation}
    \begin{aligned}
    &M^*=F_{X_2}^{-1}\left (\frac{\pi_h-\pi_m}{\pi_h-\pi_\ell}\right),\\
    &C^*=F_{X_1+X_2|X_2>M^*}^{-1}\left(\frac{\pi_m-\pi_\ell-\pi_s}{\pi_m-\pi_\ell}\right).
\end{aligned}
\end{equation}
\vspace{-1cm}
\subsection{General Single Peaked ToU without Sharing}
\vspace{-0.2cm}
Note that the $(M,C)$ policy manifests that inserting a partial peak period sophisticates the storage control in two ways: introducing additional randomness to the capacity decision and additional decision problem of discharging strategy in partial peak period. In a general single peaked ToU scheme, there are more than one partial peak periods, which will significantly increase the complexity of the storage control problem.

\begin{figure}[t]
    \centering
    \includegraphics[width=2.5in]{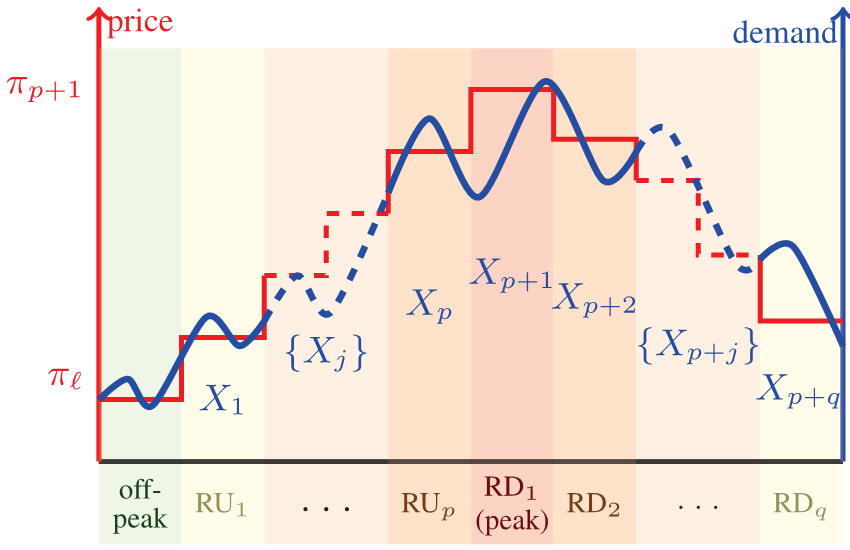}
    \caption{Single peaked ToU pricing scheme.\vspace{-0.5cm}}
    \label{fig:single_ToU}
\end{figure}

Figure \ref{fig:single_ToU} plots the general single peaked ToU scheme. The scheme consists of an off peak period, followed by $p$ ramp-up periods between off peak (denoted by $\text{RU}_j$, $j=1,\cdot\cdot\cdot,p$) and peak, followed by $q$  ramp-down periods between peak and off peak (including the peak, denoted by $\text{RD}_j$, $j=1,\cdot\cdot\cdot,q$). {$\text{RU}_j$ ($\text{RD}_j$) is also referred to as $j^{th}$ (${(p+j)}^{th}$) period of the day for brief statement. We denote the electricity rate in off peak period by $\pi_\ell$ \textcent/kWh, and the rate in $\tau^{th}$ period by $\pi_\tau$ \textcent/kWh.}

{To minimize each firm's electricity bill, the optimal} decisions for storage investment and storage control depend on the distributions of firm's random demand in each period. {We denote firm $i$'s demand in $\tau^{th}$ period by $X^{i}_\tau$.} In the subsequent analysis for standalone firm's decision making, when there is no confusion, we will drop the superscript $i$ for convenience.

We follow the $(M,C)$ type control policy for general single peaked ToU pricing. Specifically, the control policy for a firm can be decomposed as follows:
\begin{itemize}
    \item \emph{Investment Decision:} To minimize its electricity cost, the firm will invest $C$ kWh of storage.
    \item \emph{Initialization in Off Peak:} Each day, the firm will fully charge the storage system during off peak.
    \item \emph{Operation in \text{RU}s:} In $\text{RU}_j$, use the energy in the storage while reserving at least $M_j$ for future use.
    \item \emph{Operation in \text{RD}s:} In $\text{RD}_j$, use the energy in the storage as much as possible, purchase the unmet demand from the grid.
\end{itemize}
For the $j^{th}$ ramp-up period $\text{RU}_j$, we define a function $MR_j(M_j)$ as the marginal revenue for reserving $M_j$. Specifically,
\begin{equation}
    MR_j(M_j)\!\! =\!\!\!\sum_{k=j+1}^{p+q}\!\!\!(\pi_k\!-\!\pi_\ell)P_j^k(M_j,M_{j+1}^*,\cdot\cdot\cdot,M_p^*),
    \label{reservation}
\end{equation}
where $P_j^k(M_j,M_{j\!+\!1}^*,\cdot\cdot\cdot,M_p^*)$ is the probability that since $\text{RU}_j$, $k^{th}$ ($k\!>\!j$) period is the first time that the firm {needs to purchase} electricity from the grid, given optimal reservations {for following periods} $M_{j+1}^*,\cdot\cdot\cdot,M_p^*$. {
The marginal revenue measures the expected additional profit that one more unit energy reservation may bring to the firm. This one more unit reservation brings profit exactly at the first time that the firm needs to purchase energy from the grid, which is the economic intuition of Eq. (\ref{reservation}). In order to guarantee the uniqueness of optimal reservation, we make the following technical assumption:

\noindent\emph{Assumption $7$:} For each firm's demand in each non off peak period (denoted by $X$), its probability density function $f_X(x)$ is differentiable, and $f_X(x)\!>\!0$ if $x\geq0$.

With this assumption, we can show:
\begin{lemma}
If Assumption 7 is satisfied, then $MR_j(M_j)$ is monotone and the optimal reservation is unique.
\end{lemma}

The optimal control policy in $\text{RU}_j$ balances the opportunity costs of energy purchasing at current time slot and the expected marginal revenue of reserving for future at the crossing point:
\begin{equation}
    \pi_j-\pi_\ell =MR_j(M_j^*).
\end{equation}}\\
\noindent{The} associated optimal investment is {the capacity $C^*$} balancing the amortized fixed cost and the marginal benefit of investment:
\begin{align}
    \pi_s &=\sum\nolimits_{k=1}^{p+q}(\pi_k-\pi_\ell)P_0^k(C^*,M_1^*,\cdot\cdot\cdot,M_p^*),
    \label{invest}
\end{align}
where $P_0^k(C,M_1^*,\cdot\cdot\cdot,M_p^*)$ is the probability that since the off peak, $k^{th}$ period is the first time that the firm need purchase electricity from the grid, given {storage investment $C$ and the} optimal reservations $M_1^*,\cdot\cdot\cdot,M_p^*$. %The Left-Hand-Side (LHS) of (\ref{invest}) is the marginal cost of investment while the Right-Hand-Side (RHS) is the expected marginal benefit.

The above {policy} builds the foundation for analyzing {storage sharing} market. {It} is used to calculate the social optimal strategy benchmarking the efficiency of the sharing mechanism, and to examine the firm's strategy in the sharing economy. {Note that in both RU and RD periods, the optimal storage control action for each individual firm is to discharge the storage (while reserving certain energy during RU periods). This greedy control action will change dramatically when sharing is allowed.}
\vspace{-0.1cm}
\section{Storage Sharing and Investment}
\label{sharingmechanism}
\vspace{-0.1cm}

{It warrants designing a sequence of mechanisms to coordinate firms' price-arbitrage decisions. Those mechanisms respectively induce an aggregated investment decision and a set of storage sharing market. We assume that an aggregator coordinates the sharing market. We want to emphasize that in our subsequent game theoretic analysis, we do not make price taker assumption. In fact, we explicitly consider firm's competition, which leads to the efficient sharing price.
}

Two coupled games respectively arrange the mechanism of the storage investment and the associated storage sharing rules. The firms first collectively determine the {storage} capacity according to their cumulative net benefit from the storage utilization. Then, the firms {interact with the aggregator for sharing the storage to arbitrage against the ToU price in each period. Next, we severally describe the two games.}
\vspace{-0.3cm}
\subsection{Aggregator-Firm Interaction Game}
%and the firms;
%\noindent \emph{Strategies:} In $\tau^{th}$ period, the aggregator decide the sharing price $\pi^{\tau}$ while each firm $i$ correspondingly decides its charging or discharging amount;
The sharing market {can be formulated as a Stackelberg game economically enabling energy trading between the firms. In this game, a non-for-profit aggregator pursues for minimizing the total electricity cost in the industrial park. The aggregator acts as the game leader and announces the sharing price in every period while the firms respond to the price as followers. Thus, we can formally define the Aggregator-Firm Interaction Game (AFIG) as follows:

\emph{Aggregator-Firm Interaction Game (AFIG)}

\emph{Players: }The aggregator is the leader, and the firms are the followers.

\emph{Strategy Spaces: }The aggregator sets the sharing prices $\pi_\tau^a \in \mathbb{R}^+$ at each period $\tau,\quad\! \tau\!=\!1,\cdots,p+q$; the firms respond to the prices by deciding their energy procurement or supply $D^i_\tau \in \mathbb{R},\quad\! \tau\!=\!1,\cdots,p+q$ (positive $D^i_\tau$ implies energy procurement for firm $i$ at period $\tau$ while negative $D^i_\tau$ implies energy supply to the sharing market).

\emph{Utilities: }The non-for-profit aggregator seeks to minimize the total energy cost in the industrial park with its own budget balance guaranteed, while each firm seeks to minimize its own electricity bill.

To better characterize each firm's utility function, we formulate the following optimization problem for each firm $i$:
\begin{align}
    \min_{D^i_1,\cdots,D^i_{p+q}}\  &\mathcal{J}^i_{\tau}(D^i_1,\cdots,D^i_{p+q}|\pi^a_1,\cdots,\pi^a_{p+q})=\nonumber\\
    &\underbrace{\mathbb{E} \Big\{\sum\nolimits_{j=1}^{p+q} \pi^a_j D^i_j\Big\}}_{\text{\tiny cost in non off peak periods}}+\underbrace{\pi_\ell \mathbb{E}\{D_0^i\}}_{\text{recharge during off peak}},
    \label{equ:FMsharedecision}
\end{align}

The $D^i_\tau$'s are used to meet firm $i$'s demand at each period $\tau$ and simultaneously guarantee the storage temporal constraints. We choose not to explicitly write out the constraints, since we have derived the optimal $(M,C)$ policy for single firm. The aggregator can simple seek to set $\pi^a_1,\cdots,\pi^a_{p+q}$, such that the firms' aggregate behavior follows the optimal $(M,C)$ policy.

Note that, in AFIG, we assume that the firms have already purchased certain amount capacity of storage systems, i.e., $C_1,\cdots,C_n$. Hence, the equilibrium prices $\pi^a_1,\cdots,\pi^a_{p+q}$ are dependent on the storage capacities. Next, we introduce the Capacity Decision Game for firms to make the optimal storage investment decision for itself.
}

{
\subsection{Capacity Decision Game}
In the Capacity Decision Game (CDG), the firms compete for storage investment. Each firm's decision balances between investing more capacity at the beginning and exposing to more risks in the future. The profit of investing more storage comes from: the higher cost saving due to avoiding the energy procurement during the high-price periods as well as the larger profit from selling the energy stored in the storage.
We formally introduce the CDG as follows:

\emph{Capacity Decision Game (CDG)}

\emph{Players: }All the firms in the industrial park.

\emph{Strategy Spaces: }Storage investment decision $C_i \in \mathbb{R}^+$ for each firm $i$.

\emph{Utility: }Each firm seeks to minimize its expected daily cost $\mathcal{I}_i$, i.e.,
\begin{align}
    \min_{C_i}\ &\mathcal{I}_i(C_i,C_{-i})=\nonumber\\
    &\underbrace{\pi_s C_i}_{\text{invest cost}}
    \!\!+\!\!\underbrace{\sum\nolimits_{j=1}^{p+q} \mathbb{E}\{\pi_j^a D_j^i\}}_{\text{cost in non off peak periods}}
    \!\!+\underbrace{\pi_\ell \mathbb{E}\{D_0^i\}.}_{\text{recharge cost}}\label{equ:FMCDG}
\end{align}
Here, $C_{-i}=\sum_{j\not=i} C_j$ refers to the sum of all other firms' capacities. More precisely, note that $\pi_j^a$'s are determined by $C_{i}$ and $C_{-i}$, which naturally leads to the game formulation. $D_j^i$ and $D_0^i$ are actually related to $C_i$.
}
\begin{figure}[t]
    \centering
    \includegraphics[width=3.1in]{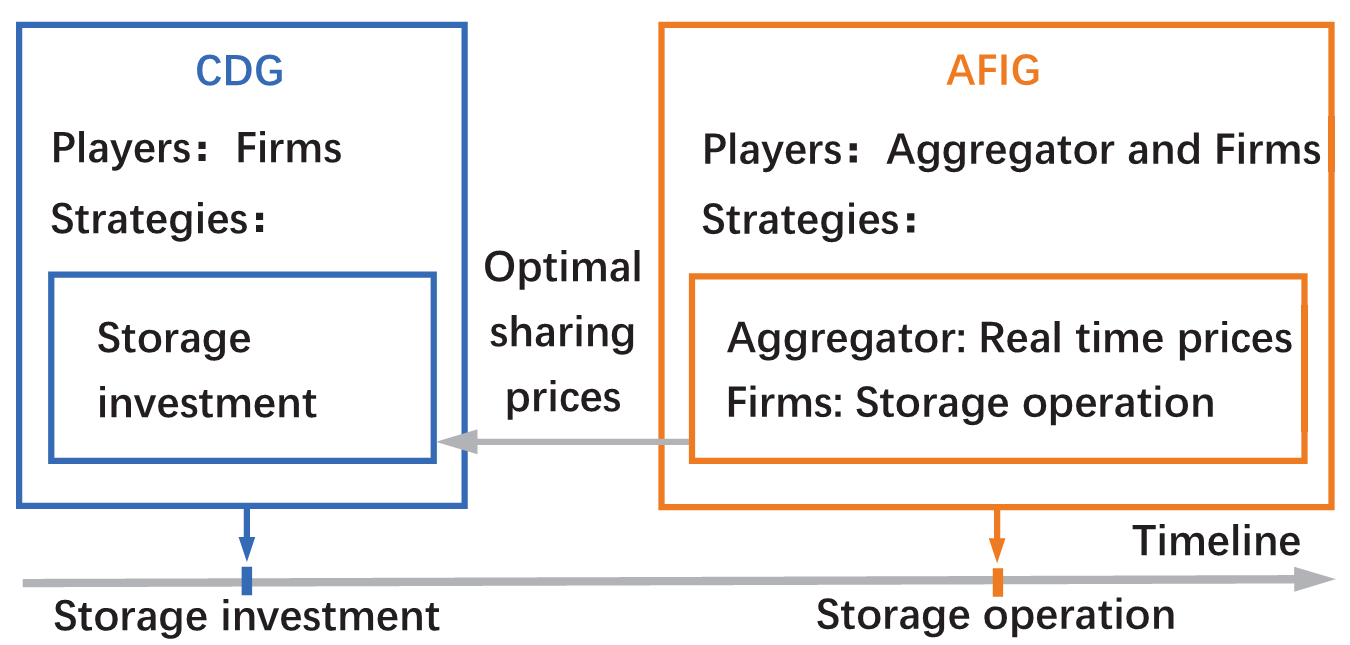}
    \caption{The coupling between two games: CDG and AFIG.}
    \label{fig:coupling}
\end{figure}
{

\noindent \textbf{Remark:} The outcome of CDG heavily depends on the aggregator's pricing scheme in AFIG. This is how the two games are coupled together. We visualize the coupling in Fig. \ref{fig:coupling} from two perspectives: the time sequence and the analytical sequence. Specifically, in the timeline, the firms first determine the capacity (the decision of CDG), and then the aggregator arranges the AFIG each day. However, to analyze the optimal decisions to the two games, we need to solve them in backward. We first analyze the {N.E.} (especially the aggregator's sharing prices) in the AFIG, and then, based on {this} information, we can understand the optimal storage investment decision for each firm in the CDG.

To ease the understanding of the sequential decision making, we first analyze the coupling dynamics through simple cases (three-tier ToU) and then generalize the results to general single peaked ToU.
}
\vspace{-0.3cm}
\section{Market Analysis for Three-Tier ToU}
\vspace{-0.2cm}
\label{3T}
This section discusses the firms' strategies and market equilibria in the three-tier ToU, which reveals the intuition of our proposed pricing mechanism, and the corresponding firm strategies as well as associated system impacts. There are two basic scenarios: the ramp-up scheme (RUS) and the ramp-down scheme (RDS). We separately discuss these two basic scenarios. 
{We first notice that the following two facts are true in both the RUS and RDS.

\begin{fact}\label{fact:1sthr}
    Firms' optimal strategy in the off-peak period is to fully recharge the storage without sharing their energy.
\end{fact}
\begin{fact}    
    Firms only reserve energy during RU periods.
\end{fact}
These two facts allow us to narrow our focus on the firms' strategies during the peak and partial peak periods, and their associated market dynamics.}

%split the whole day into three parts and separately analyze the equilibrium and associated player's optimal strategies in each part. The three parts are the first hour, the ramp-up period, and the ramp-down period.

\vspace{-0.3cm}
\subsection{Ramp-Down Scheme}
\vspace{-0.2cm}
In the RDS (as shown in Fig. \ref{fig:ToU_3} (b)), we model firm $i$'s peak (partial peak) demand as random variable {$X^i_{1'}$ ($X^i_{2'}$)}, and define the total demand during peak and partial peak periods by {$X^c_{1'}$} and {$X^c_{2'}$}, i.e.,
{
\begin{align*}
    {X_{1'}^c}=\sum\nolimits_i {X^i_{1'}}; \quad {X_{2'}^c}=\sum\nolimits_i {X^i_{2'}}.
\end{align*}
}\\
Observing a greedy type policy is optimal, the follow fact dictates the {N.E.} during the partial peak period when the $i^{th}$ firm has $u^i$ units of electricity in the storage {at the beginning of partial peak period} and has to consume {$x^i_{2'}$} units of electricity ({the realization of $X^i_{2'}$}) in the partial peak period.

\begin{fact}\label{fact:RDSPartialEq}
The optimal electricity sharing price {for maximal social welfare} in partial peak should be set as
\begin{align}
\pi_{partial}^{a*}=
\left\{
\begin{array}{rcl}
{\pi_{2'}},&\text{if } {x_{2'}^c }\geq u^c, \\
\pi_\ell,&\text{if } {{x_{2'}^c }< u^c},
\end{array} \right.
\end{align}
where ${x_{2'}^c }=\sum_i {x_{2'}^i }$, denoted the realization of $X_{2'}^c$; and $u^c=\sum_i u^i$.
%$\hfill\blacksquare$
\end{fact}
This fact is simply the result of competition among firms in different market conditions.

In peak period, we propose the optimal pricing strategy {for the aggregator to minimize the total cost of all firms} when the {realized} demand is {$x_{1'}^i$} for firm $i$: 
\newtheorem{mechanism}{Mechanism}
\begin{mechanism}
The optimal electricity sharing price in peak period {of the aggregator} $\pi_{peak}^{a*}$ should be set as:
\begin{align}
    \pi_{peak}^{a*}=\left\{\begin{array}{ccc}
{\pi_{1'} }, \text{if } {x_{1'}^c} > C_c,\\
\pi', \text{if } {x_{1'}^c} <C_c,
\end{array} \right.\nonumber
\end{align}
\label{price}
where $\pi'=\pi_\ell+({\pi_{2'}}-\pi_\ell)\text{Pr}({x_{2'}^c }>C_c-{x_{1'}^c})$, ${x_{1'}^c}=\sum_{i\in \mathcal{N}}{x^i_{1'}}$ and $C_c=\sum_i C_i$.
\end{mechanism}
The mechanism is derived from the aggregator's system-cost minimization problem thus guarantees the optimal market equilibrium {for social welfare}, yielding the following lemma: 
\begin{lemma}
    {The aggregator adopts pricing scheme in Mechanism 1 will lead the firms to make storage operations which supports social welfare (in terms of minimizing the total cost), while achieves aggregator's budget balance (i.e., zero cost for the aggregator).}
\label{optandbal}
\end{lemma}
\begin{itemize}
\item When the total {realized} demand exceeds the total amount of stored energy ({$x_{1'}^c > C_c$}), each firm will choose to {greedily} use up its storage since its own cost monotonically increases with its reservation $u^i$:
\begin{align}
    \frac{\partial \mathcal{J}_i}{\partial u^i}&={\pi_{1'} }-\pi_\ell-({\pi_{2'} }-\pi_\ell)\text{Pr}({x_{2'}^c }>u^c) \nonumber\\
    &>{\pi_{1'} }-{\pi_{2'} }>0.
    \label{derivativeicase1}
\end{align}
%The total demand in the sharing market is $\sum_i {(x_1^i-C_i)}^+$, and the supply has no limit because they can buy electricity from the utility at the sharing price.
\item When the stored energy is sufficient to satisfy {realized} the demand (${x_{1'}^c } < C_c$), we have
\begin{align}
    \pi_{peak}^{a*}&=\pi_\ell+({\pi_{2'} }-\pi_\ell)\text{Pr}({x_{2'}^c }>C_c-{x_{1'}^c })\nonumber\\
    &=\mathbb{E}\{\pi_{partial}^{a*}|{x_{1'}^c }\}.
\end{align}
This yields that the first order optimality condition automatically holds:
\begin{align}
    \frac{\partial \mathcal{J}_i}{\partial u^i}=\pi_{peak}^{a*}-\mathbb{E}\{\pi_{partial}^{a*}|{x_{1'}^c }\}=0.
    \label{derivativecase2}
\end{align}
That is, {the firms have no incentive to make any energy reservation in this case. Together with Eq. (\ref{derivativeicase1}), Mechanism \ref{price} indeed leads to the optimal greedy policy.}
\end{itemize} 
The firms' total expected profit from owning the storage, which is framed by the mechanism organizing the sharing market, shapes their storage-investment decisions in the CDG. Each firm balances the cost and benefit from owning the storage by minimizing the total cost $\mathcal{I}_i(C_i,C_{-i})$, including the investment cost and the total expected cost from purchasing electricity.
\begin{align}
    \mathcal{I}_i(C_i,&C_{-i})=\pi_s C_i+\mathbb{E}\{\pi_{peak}^{a*} D^i_{peak}\} \nonumber\\
    &+\mathbb{E}\{\pi_{partial}^{a*} D^i_{partial}\}+\pi_\ell \mathbb{E}\{D^i_{off}\},
    \label{equ202}
\end{align}
where $C_{-i}=\{C_j|j \neq i\}$ implicitly influences the sharing prices in peak and partial peak periods, and $D^i_{peak}$, $D^i_{partial}$ and $D^i_{off}$ are firm $i$'s electricity deficit during the peak, partial peak and off peak period respectively. The {N.E.} of the CDG is summarized in the following theorem.
\begin{theorem}
If CDG admits a N.E., then it is unique and can be characterized by
\begin{align}
    C_i^*&=\lambda_1\mathbb{E}[{X^i_{1'}}|{X_{1'}^c}=C_c^*] \nonumber\\
    &+\lambda_2\mathbb{E}[{X^i_{1'}}+{X^i_{2'}}|{X_{1'}^c}+{X_{2'}^c}=C_c^*],\forall i,
\end{align}
where
\begin{align}
     \lambda_1=\frac{({\pi_{1'} }-{\pi_{2'} })f_{{X_{1'}^c}}(C_c^*)}{({\pi_{1'} }-{\pi_{2'} })f_{{X_{1'}^c}}(C_c^*)+({\pi_{2'} }-\pi_\ell)f_{{X_{1'}^c}+{X_{2'}^c}}(C_c^*)},\nonumber \\
     \lambda_2=\frac{({\pi_{2'} }-\pi_\ell)f_{{X_{1'}^c}+{X_{2'}^c}}(C_c^*)}{({\pi_{1'} }-{\pi_{2'} })f_{{X_{1'}^c}}(C_c^*)+({\pi_{2'} }-\pi_\ell)f_{{X_{1'}^c}+{X_{2'}^c}}(C_c^*)},\nonumber
\end{align}
and $C_c^*$ is the unique solution to Eq. (\ref{equ:rdinvest})
\begin{align}
    \pi_s&=({\pi_{1'} }-\pi_\ell)\text{Pr}({X_{1'}^c}> C_c)\nonumber\\
    &+({\pi_{2'} }-\pi_\ell)\text{Pr}({X_{1'}^c}<C_c,{X_{1'}^c}+{X_{2'}^c}> C_c).
    \label{equ:rdinvest}
\end{align}
\end{theorem}
%\red{Discuss how the ToU rates influence the sharing decision and benefit as well as the investment [Simulation].}
However, CDG may not always admit a N.E. in that the cost function $\mathcal{I}_i(C_i,C_{-i})$ could be non-convex with respect to $C_i$. We need the following technical alignment conditions to guarantee the existence of the {N.E.}:
%By setting $Y_{2}^c\equiv 0$ (the demands in \text{RD} vanish), a counterexample could be got from the counterexample in \cite{8025410} as mentioned before. Here we give a group of sufficient conditions for the existence of the equilibrium.
%\newtheorem{theorem}{Theorem}

\begin{theorem}
If the following two conditions hold:
\begin{align}
    & \frac{\partial \mathbb{E}[{X_{1'}^i}|{X_{1'}^c}=r]}{\partial r} \geq 0,\ \forall i,
    \label{cond12}\\
    & \frac{\partial \mathbb{E}[{X^i_{1'}}+{X^i_{2'}}|{X_{1'}^c}+{X_{2'}^c}=c]}{\partial c} \geq 0, \ \forall i,
    \label{cond22}
\end{align}
then CDG for RDS always admits a unique N.E..
\label{tech1}
\end{theorem}
%The rigorous proof can be found in Appendix XX.
{
\noindent\textbf{Remark:} The two technical alignment conditions merely require \emph{on expectation}, the energy consumption for each individual firm should align with the trend of total load. This can be guaranteed by checking the pairwise correlations. Intuitively speaking, as long as there is only a small portion of negative pairwise correlations, the alignment conditions automatically hold. We have observed the distribution of correlation for the residential loads under the SCE Grandfathered Rate Plans \cite{SCEToU}. The residential loads in Austin from Pecan street \cite{Pecan}, only produce $2.3\%$ negative pairwise correlations in peak period and $2.0\%$ for the total demands in peak and the following partial peak periods.
%The commercial building loads, with inherent pattern, produce even less negative pairwise correlations (only $7.8\%$ in terms of demands in peak period and $8.7\%$ in terms of the total demands in peak and the following partial peak demands).
}

The above analysis presents how firms' demands and ToU designs determine sharing market design as well as the storage investment.

\subsection{Ramp-Up Scheme (RUS)}
In contrast to the {greedy} strategy in RDS, the firms have an extra choice that is to reserve electricity for future in addition to sharing.  Since the analysis is almost identical, we will only present necessary intuition for {the analysis in} RUS. We denote firm $i$'s random partial peak and peak demand by $X_1^i$ and $X_2^i$, respectively, {and denote its realized partial and peak demand by $x_1^i$ and $x_2^i$.}
\begin{mechanism}
The {optimal pricing scheme for the aggregator} in RUS is to set $\pi_{peak}^{a}$ and $\pi_{partial}^{a}$ as follows:
\begin{align} 
\pi_{peak}^{a*}=
\left\{
\begin{array}{rcl}
\pi_2,&\text{if } x_2^c > u^c, \\
\pi_\ell,&\text{if } {x_2^c < u^c},
\end{array} \right.
\end{align}
\begin{align}
    \pi_{partial}^{a*}=\left\{
\begin{array}{ccc}
 \pi_1, \text{if } x_1^c> C_c-M, \\
\pi'', \text{if } x_1^c< C_c-M,
\end{array} \right.
\end{align}
\label{price4}\\
where $x_2^c=\sum_{i} x_2^i$, $u^c=\sum_{i } u^i$ and $\pi''=\pi_\ell+(\pi_2-\pi_\ell)\text{Pr}(X_2^c > C_c-x_1^c)$, {and
\begin{equation}
    M=F_{X_2^c}^{-1}\left (\frac{\pi_2-\pi_1}{\pi_2-\pi_\ell}\right).
\end{equation}
}
\end{mechanism}
We can show the optimality of Mechanism \ref{price4} through the following lemma:
\begin{lemma}
    {The aggregator adopts pricing scheme in Mechanism 2 will lead the firms to make storage operations which supports social welfare (in terms of minimizing the total cost), while achieves aggregator's budget balance (i.e., zero cost for the aggregator).}
\label{optandbal1}
\end{lemma}

We would also like to point out that in RUS, there exists another CDG and this time the technical alignment condition to guarantee {the} existence of N.E. is different:
\begin{theorem}
If for each firm $i$,
\begin{align}
    \frac{\partial \mathbb{E}[X_1^i+X_2^i|X_1^c+X_2^c=r,X_1^c <r- M]}{\partial r} \geq 0,
    \label{cond2}
\end{align}
Then, the CDG for RUS admits a unique N.E..
\label{tech2}
\end{theorem}
{
\noindent\textbf{Remark:} Again, this technical alignment condition is not hard to meet. It also requires \emph{on expectation}, the energy consumption for each individual firm should align with the trend of total load.
}
\vspace{-0.3cm}
\section{General Single Peaked ToU Scenario}
\vspace{-0.2cm}
\label{General}
Having the intuition from analyzing three-tier ToU schemes, we know that the whole decision process couples with two games: AFIG and CDG. In this section, we analyze the sharing mechanism for single peaked ToU scheme. Again, the firms will follow the agggregator's price signal to make their sharing decisions to minimize their own cost in the non off peak periods, and make storage sizing decisions according to CDG's {N.E.}. However, the temporal coupling brings huge difficulties.

For $\tau^{th}$ non off peak period, $u_{\tau-1}^c$ refers to the total energy left in the storage at the beginning of the period; $x_\tau^c$ refers to the realized total demand in the period; $P_\tau^k(x_\tau^c,u_{\tau-1}^c)$ refers to the probability that since {$\tau^{th}$ non off peak period, $k^{th}$ period ($k\!>\!\tau$) is the first time that the collective need purchase electricity from grid, based on the realized demand $x_\tau^c$ and the initial reservation $u_{\tau-1}^c$. 
Based on these quantities, we can use the same intuition of Eq. (\ref{reservation}) to obtain the expected marginal profit for reserving energy at each time slot. On the other hand, the marginal cost for reserving energy at time $\tau$ is simply $\pi_\tau^a$, the sharing price that the energy set in $\tau^{th}$ period. And we seek to design the optimal prices ($\pi_\tau^{a*}$'s) {of the aggregator}. 
%$\pi_{\tau}^{a}$ refers to the sharing price that the aggregator should set in $\tau^{th}$ period. We seek to obtain the optimal price $\pi_{\tau}^{a*}$'s.

For $j^{th}$ non off peak period, $j=p\!+\!q,\cdot\cdot\cdot,1$, we propose that the sharing price $\pi_{j}^{a}$ should satisfy
    \begin{align}
        \pi_{j}^{a*}\!=\!\pi_\ell\!+\!\!\sum_{k=j}^{p+q} (\pi_k\!-\!\pi_\ell) P_j^k(x_{j}^c, u_{j-1}^c).
        \label{mechanism3}
    \end{align}
    Notice that if realized demand ($x_{j}^c$) is more than total energy left  ($u_{j-1}^c$), then the collective need purchase electricity from grid in period this period; i.e., $P_j^j=1$, yielding $\pi_{j}^{a*}=\pi_{j}$, which coincides with intuition.

}

This sharing mechanism leads to another CDG: if the game admits a N.E., it is unique and given by
\begin{equation}
    \begin{aligned}
    C_{i}^*\! =\!&\!\sum_{l=1}^{q} \rho_l(C_c^*)\mathbb{E}\Bigg[ \sum_{k=1}^{p+l}\! X^i_k\Bigg|\sum_{k=1}^{p+{l}} X_k^c=\!C_c^*,\\
    &\ \ \ \ \sum_{k=1}^n X_k^c<C_c^*-M_n^*,1\leq n\leq p\Bigg],
        \label{indinvest}
    \end{aligned}
\end{equation}
where
\begin{align*}
    &\rho_l(C_c^*)\!=\!\frac{(\pi_{p+l}-\pi_{p+l+1})f_{\{\sum_{k=1}^{p+l} X_{k}^c\}}(C_c^*)}{\sum_{m=1}^{q} (\pi_{p+m}-\pi_{p+m+1})f_{ \{\sum_{k=1}^{p+m} X_{k}^c\}}(C_c^*)},
\end{align*}
and $C_c^*$ is the unique solution of
\begin{equation}
    \pi_s \!=\!\!\sum\nolimits_{k=1}^{p+q}(\pi_k\!-\!\pi_\ell)P_0^k(C_c),
    \label{Cc}
\end{equation}
where $\pi_{p+q+1}\!\!=\!\!\pi_\ell$, $P_0^k(C_c)$ refers to the probability of that since off peak, $k^{th}$ non off peak period is the first time that the collective need purchase electricity grid, based on investment $C_c$. To ensure the existence of N.E., we need the following technical alignment conditions:
\begin{theorem}
If $\forall \text{RD}_j,\ j:1\leq j\leq q $,
\begin{align}
    \frac{\partial G_{j}^i(r)}{\partial r}\geq 0,\ \forall i
    \label{align}
\end{align}
where 
\begin{align*}
    G_j^i(r)=&\mathbb{E}\Bigg [
     \sum_{k=1}^{p+j} X_k^i\Bigg|\sum_{k=1}^{p+j} X_k^c\!=\!r,\\
    &\qquad\sum_{k=1}^n X_k^c<r-M_n^*,1\leq n\leq p
    \Bigg],
\end{align*}
then the {N.E.} of CDG in the single peaked multi-tier ToU pricing exists.
\end{theorem}
\noindent \textbf{Remark:} {The N.E.'s of AFIG and CDG support social welfare for all firms.} {By the definition of equilibrium, they can automatically ensure individual rationality.} In fact, they enjoy additional properties, as summarized in the following proposition:
\begin{proposition}
\label{prop}
The equilibria of the sharing markets and CDG in single peaked multi-tier ToU pricing scheme enjoy the following properties: 

(a) \emph{Coalitional stability}: No subset of firms are better off by defecting to form their own coalition.

(b) \emph{No pure-strategy play}: If firm $i$ has no demand in all periods other than the off peak, it will not invest in storage. Hence, the aggregator can be in a neutral position to coordinate the market.
\end{proposition}

\vspace{-0.2cm}
\section{Simulation Studies}
\label{Simulation}
\vspace{-0.2cm}
To evaluate the performance of our control policy and sharing mechanism, we use real household profiles (in summer of 2016) from Austin in Pecan Street \cite{Pecan}. We compare our control policy with a simple policy where the decisions for all the firms across the day are decoupled. That is, the firms won't reserve energy for future use. They will focus on the arbitrage (and sharing) opportunities between off peak period and other periods. In a single peaked pricing scheme with $k$ periods, this simple policy is the same as solving $k\!-\! 1$ 2-tier (one non off peak and off peak) storage sharing problems. We refer to this simple policy as ``2-tier division''.
\begin{figure}[t]
\centering
\begin{tikzpicture}[xscale=1, yscale = 0.3]
%increasing mode
\draw[very thick,->] (0,0) -- coordinate (x axis mid) (5,0);
%\node[align=right] at (2,-1) {};
\node at (0.4,5.75) {price};
\draw[very thick,->] (0,0) -- coordinate (y axis mid) (0,6);
\draw [very thick,blue] (0,1.3) --  (1.6,1.3)--(1.6,2.8)--(2.4,2.8)--(2.4,5.2)--(4,5.2)--(4,2.8)--(4.4,2.8)--(4.4,1.3)--(4.8,1.3);
\draw [draw=none, fill=green, fill opacity = 0.2] (0,0) rectangle (1.6,1.3);
\draw [draw=none, fill=orange, fill opacity = 0.2] (1.6,0) rectangle (2.4,2.8);
\draw [draw=none, fill=red, fill opacity = 0.2] (2.4,0) rectangle (4,5.2);
\draw [draw=none, fill=orange, fill opacity = 0.2] (4,0) rectangle (4.4,2.8);
\draw [draw=none, fill=green, fill opacity = 0.2] (4.4,0) rectangle (4.8,1.3);
\node[anchor=east] at (0,1.3) {\small 13\textcent};
\node[anchor=east] at (0,2.8) {\small 28\textcent};
\node[anchor=east] at (0,5.2) {\small 52\textcent};
\node[anchor=north, text=black] at (0,-0.03) {\scriptsize 0AM};
\node[anchor=north, text=black] at (1.6,-0.03) {\scriptsize 8AM};
\node[anchor=north, text=black] at (2.4,-0.03) {\scriptsize 2PM};
\node[anchor=north, text=black] at (3.9,-0.05) {\scriptsize 8PM};
\node[anchor=north, text=black] at (4.4,-0.03) {\scriptsize 10PM};
\node[anchor=north, text=black] at (5,-0.03) {\scriptsize 12PM};
%\draw [very thick,blue,dashed] (1.25,0) --  (1.25,1.5);
%\node[anchor=north, text=blue] at (1.25,-0.1) {\scriptsize $\sum_i {(C_i-x_1^i)}^+$};
%\node[anchor=east] at (0,1.5) {\small $\pi'$};
%\node[anchor=east] at (0,3.5) {\small $\pi_1$};
%\node[align=left,rotate=90,text=red] at (0.5,3) {\small Demand};
%\node[anchor=south,text=blue] at (1.85,3.5) {\small Supply};
%\filldraw (0.65,1.5) ellipse (1.6pt and 2.3pt);
%\draw[very thick,->] (1.35,2)--(0.72,1.55);
%\node at (1.9,2) {\small equil point};
\end{tikzpicture}
\caption{SCE Grandfathered Rate Plans ToU-D-A \cite{SCEToU}.\vspace{-0.3cm}
}
\label{fig:simulation pricing}
\end{figure}
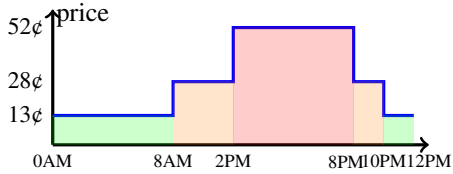

\begin{figure}[t]
    \centering
    \includegraphics[width=2.8in]{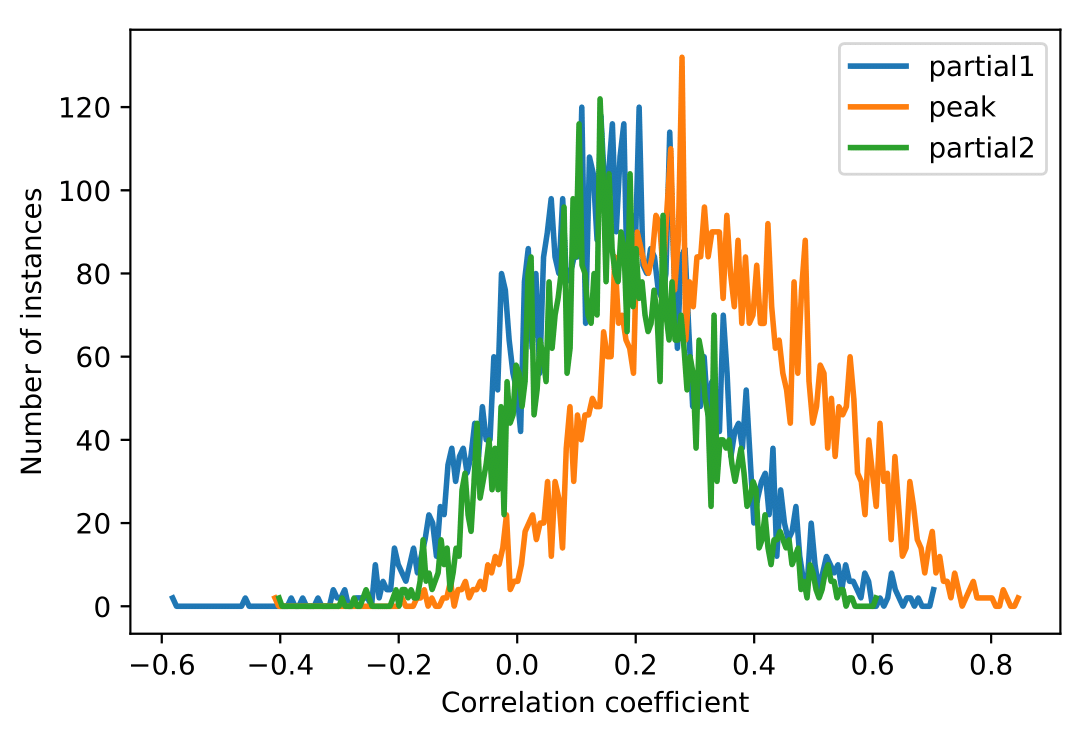}
    \caption{Diversity of partial peak and peak demands.\vspace{-0.2cm}}
    \label{diversity}
\end{figure}

We use a real 3-tier ToU pricing scheme as shown in Fig. \ref{fig:simulation pricing} for simulation. We use Tesla Power 2.0 Lithium-ion to estimate $\pi_s$: amortized over 10-year lifetime, the Tesla battery costs 14 \textcent/kWh per day \cite{Tesla}.
%To get benefit from storage sharing, the demands in partial peak and peak period must be diversified. Histograms of the pairwise correlation coefficients of real demands are shown in Fig. \ref{diversity}.

{Pairwise correlation coefficient reflects the sharing market conditions. The pairwise correlation coefficient between the consumptions of firm $a$ and firm $b$ in $j^{th}$ period can be calculated as follows:
\begin{align}
    \rho_j^{ab}\!=\!\frac{\mathbb{E}\{X_j^a X_j^b\}-\mathbb{E}\{X_j^a\}\mathbb{E}\{X_j^b\}}{\sqrt{\text{Var}\{X_j^a\}\cdot\text{Var}\{X_j^b\}}}.
\end{align}
}\\
{If most coefficients are around 1, then there will be little room for sharing. Figure \ref{diversity} plots the histogram of the pairwise correlation coefficients of the real demands in different periods. We can observe that the mean coefficient is around 0.16 while only $7\%$ of coefficients exceed $0.5$. This illustrates certain room for sharing markets in practice.}

\begin{figure}[t]
    \centering
    \includegraphics[width=2.7in]{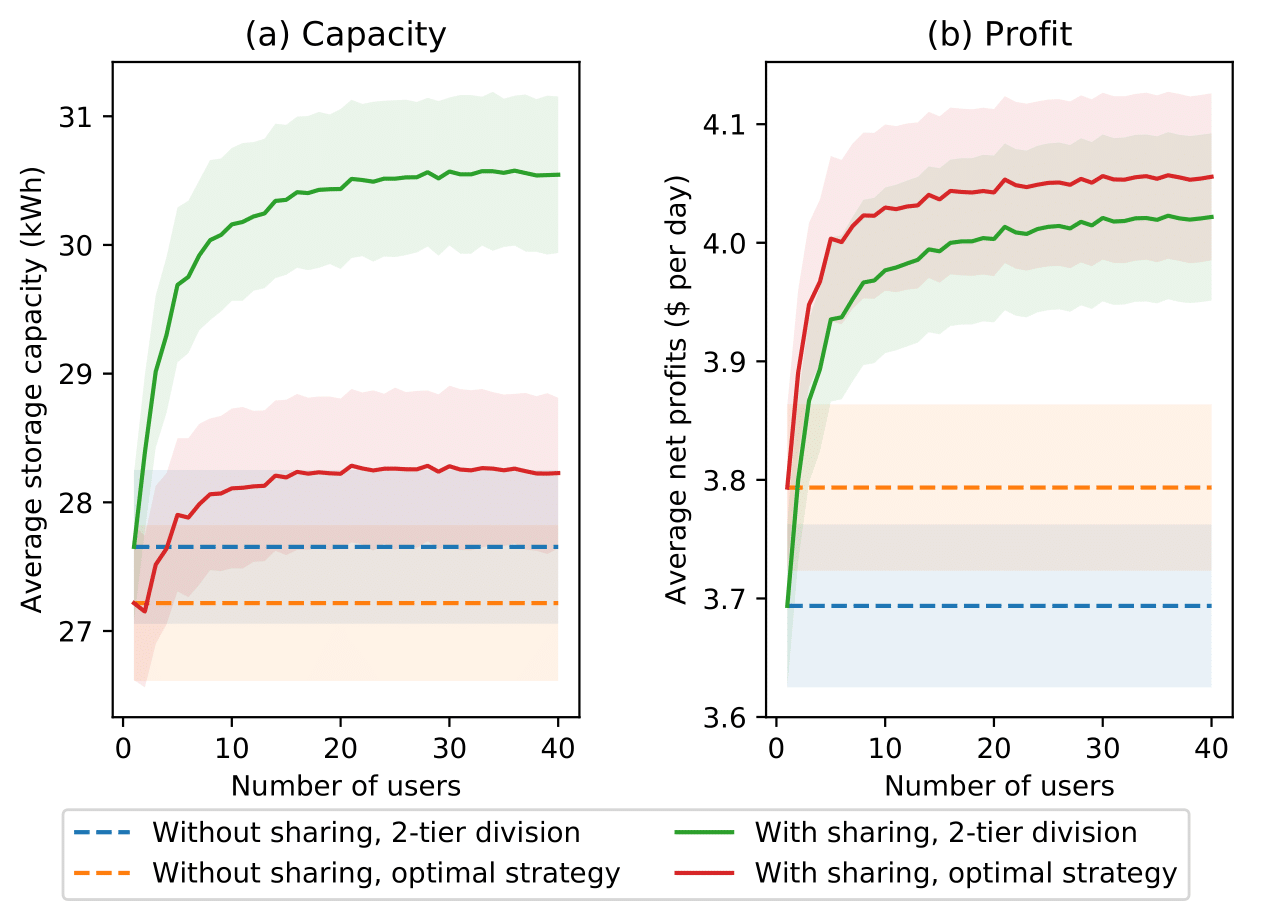}
    \caption{Capacity and profit for different mechanisms. The shaded regions in Fig. 6 reflects the interval between mean-var/20 and mean+var/20 over different community size.\vspace{-0.5cm}}
    \label{fig:simulation}
\end{figure}

Figure \ref{fig:simulation} compares the different control policies. Our proposed sharing mechanism achieves about 7 \textcent/day more profit compared with 2-tier division and about 23 \textcent/day more profit compared with no sharing scenario. If the industrial park owner strategically selects the firms with more diversified load profiles, our sharing mechanism can yield more profits.

\begin{figure}[t]
    \centering
    \includegraphics[width=3in]{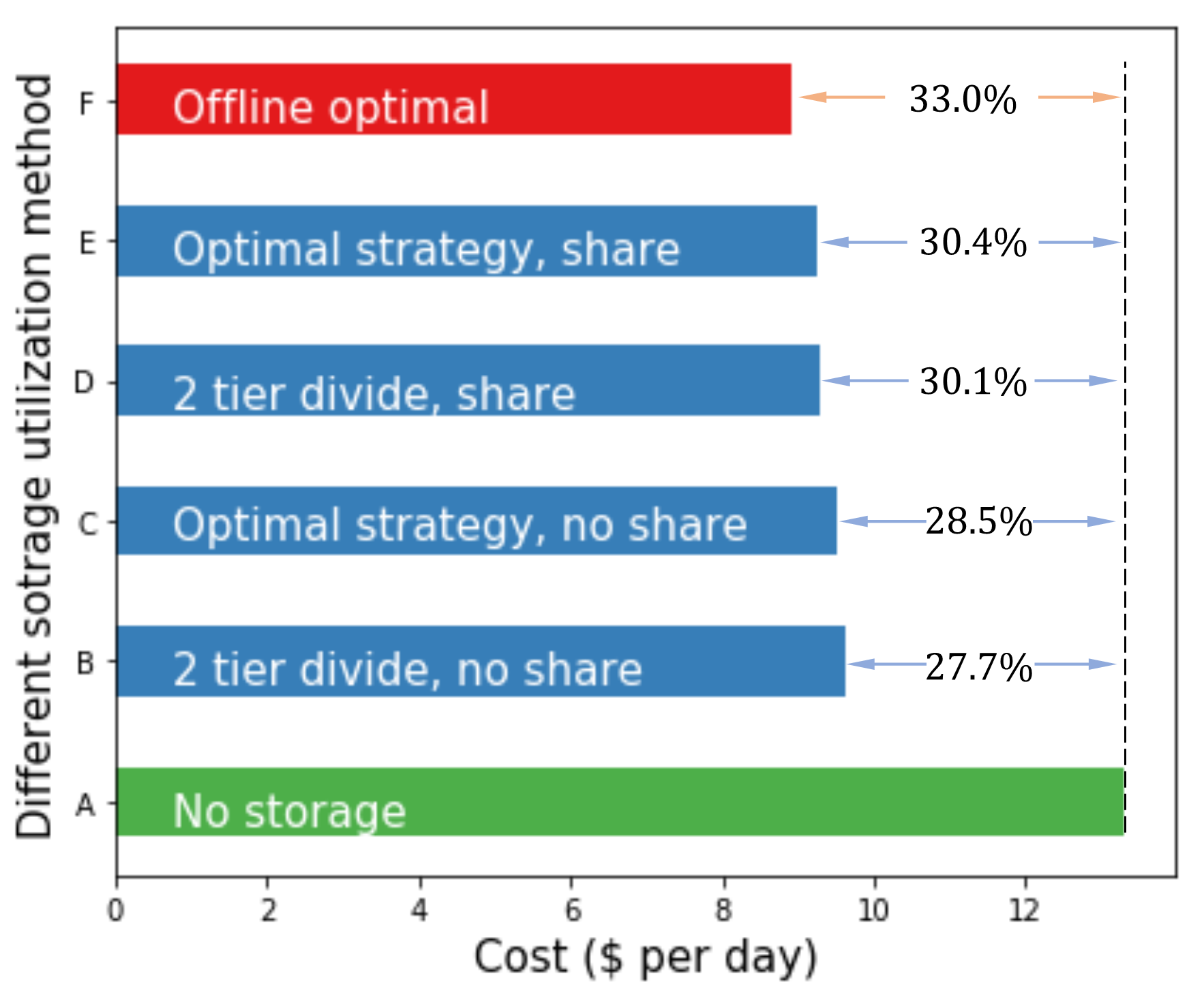}
    \caption{Cost comparison between different kinds of storage utilization.}
    \label{fig:compare}
\end{figure}

{We highlight the impact of independent demand assumption is not too restrictive in Fig. \ref{fig:compare}. The offline optimal denotes the storage operation with all the future demand information, which yields $33.0\%$ average cost saving compared to the benchmark when there is no storage in the system at all. Our proposed optimal storage sharing mechanism achieves $30.4\%$ expected cost saving. The $2.6\%$ difference is the aggregate effects of unknown future information and the independent demand assumption.
}
\vspace{-0.2cm}
\section{Conclusions and Future Work}
\label{Conclusionsection}
\vspace{-0.2cm}
We propose a sharing mechanism for electricity storage for single peaked ToU pricing scheme. We show that this sharing mechanism supports the maximal social welfare.

We are interested in understanding the sharing mechanism for the dynamic pricing schemes. The uncertainties in the electricity rates will incur challenge for theoretical analysis. It will be also interesting to consider the sharing market where the firms make their decisions sequentially instead of simultaneously. This scenario reflects the dynamics of firm's joining the industrial park.

\bibliographystyle{IEEEtran}
\bibliography{IEEEabrv,mylib}

\appendix
{
\subsection{Assumption 5 Relaxation}
\vspace{-0.2cm}
%In this subsection, we drop our assumption 5 by considering energy loss in storage operation. We assume the charging efficiency is $\eta_i$ and the discharging efficiency is $\eta_o$. Now we can rearrange our mechanism and provide the associated equilibrium results.
The imperfect charging efficiency and discharging efficiency will heavily affect the storage sharing mechanism characterization. Denote the charging efficiency by $\eta_i$ and the discharging efficiency by $\eta_o$. Below, we show how to characterize the $(M, C)$ control policy, the sharing price $\pi_j^a$'s, the optimal storage investment $c_i^*$ for each firm $i$, and the technical alignment conditions, all in terms of $\eta_i$ and $\eta_o$.

With the charging and discharging efficiency, The optimal collective reservation in $j^{th}$ non off peak period $M_j$ can be formulated as:
\begin{equation}
    \pi_j\eta_o-\pi_\ell/\eta_i\!\! =\!\!\!\sum_{k=j+1}^{p+q}\!\!\!(\pi_k\eta_o\!-\!\pi_\ell/\eta_i)P_j^k(M_j,M_{j+1}^*,\cdot\cdot\cdot,M_p^*).
\end{equation}

For $j^{th}$ non off peak period, $j=p\!+\!q,\cdot\cdot\cdot,1$, the sharing price $\pi_{j}^{a}$ should satisfy
\begin{align}
    \pi_{j}^{a*}\eta_o\!=\!\pi_\ell/\eta_i\!+\!\!\sum_{k=j}^{p+q} (\pi_k\eta_o\!-\!\pi_\ell/\eta_i) P_j^k(x_{j}^c, u_{j-1}^c).
\end{align}

This sharing mechanism again leads to a CDG: if the game admits a N.E., it is unique and given by
\begin{equation}
    \begin{aligned}
    C_{i}^*\! =\!&\!\sum_{l=1}^{q} \rho_l(C_c^*)\mathbb{E}\Bigg[ \sum_{k=1}^{p+l}\! X^i_k\Bigg|\sum_{k=1}^{p+{l}} X_k^c=\!C_c^*\eta_o,\\
    &\ \ \ \ \sum_{k=1}^n X_k^c<(C_c^*-M_n^*)\eta_o,1\leq n\leq p\Bigg],
    \end{aligned}
\end{equation}
where
\begin{align*}
    \rho_l(C_c^*)\!\!=\!\!\!\quad\!\frac{(\pi_{p+l}\!-\!\pi_{p+l+1})f_{\{\sum_{k=1}^{p+l} X_{k}^c\}}(C_c^*\eta_o)}{\sum_{m=1}^{q} (\pi_{p+m}-\pi_{p+m+1})f_{ \{\sum_{k=1}^{p+m} X_{k}^c\}}(C_c^*\eta_o)},
\end{align*}
and $C_c^*$ is the unique solution of
\begin{equation}
    \pi_s\eta_o\!=\!\!\sum\nolimits_{k=1}^{p+q}(\pi_k\eta_o\!-\!\pi_\ell/\eta_i)P_0^k(C_c),
\end{equation}
where $\pi_{p+q+1}\!\!=\!\!\pi_\ell/(\eta_i\eta_o)$, $P_0^k(C_c)$ refers to the probability of that since off peak, $k^{th}$ non off peak period is the first time that the collective need purchase electricity grid, based on investment $C_c$. 

To ensure the existence of N.E., we need the following technical alignment conditions:

\begin{proposition}
If $\forall \text{RD}_j,\ j:1\leq j\leq q $,
\begin{align}
    \frac{\partial G_{j}^i(r)}{\partial r}\geq 0,\ \forall i
\end{align}
where 
\begin{align*}
    G_j^i(r)=&\mathbb{E}\Bigg [
     \sum_{k=1}^{p+j} X_k^i\Bigg|\sum_{k=1}^{p+j} X_k^c\!=\!r\eta_o,\\
    &\qquad\sum_{k=1}^n X_k^c<(r-M_n^*)\eta_o,1\leq n\leq p
    \Bigg],
\end{align*}
then the {N.E.} of CDG in the single peaked multi-tier ToU price exists.
\end{proposition}

\vspace{-0.2cm}
\subsection{Optimality of the Control Policy}

\textbf{Remark:} We only prove those results for the general single peaked ToU in the following parts. Results for the two basic 3-tier ToU schemes are special cases of the general scenario and we only use them to obtain more intuitions in the main content.

In this subsection, we show that the arbitrage control policy proposed in Section \uppercase\expandafter{\romannumeral3}-B can bring maximal profit using backward induction. Firstly, we prove that the greedy policy is optimal in RD periods. Secondly, we prove that the threshold policy is optimal in RU periods and characterizes the best thresholds at the same time. Finally, we characterize the optimal investment decision.   

For $j^{th}$ non off peak period of the day, we denote energy left in the storage at the beginning by $u_{j-1}$, and energy left in the storage at the end by $u_{j}$, the electricity we need to purchase by $D_j$ and the realized demand in this period as $x_j$. Let $D_0$ be the electricity amount we need to purchase in off peak for recharge.

\textbf{Greedy Policy in RD Periods Is Optimal}

We first examine the trivial case in {$\textbf{RD}_{{q}}$}, the last non off peak period. It is straightforward to see that the optimal strategy is to discharge the storage as much as possible, which constructs our induction basis.

Now we prove the optimality for \textbf{the rest RD periods}. Assume that greedy policy is optimal from $\text{RD}_q$ to $\text{RD}_{j+1}$. With this hypothesis, we want to show that greedy policy is optimal in $\text{RD}_j$. 

The total cost $\mathcal{J}_{p+j}$ in $\text{RD}_j$ and following RD periods is as follows:
\begin{align}
    \mathcal{J}_{p+j}&(u_{p+j})
    =\underbrace{\pi_{p+j}(u_{p+j}+x_{p+j}-u_{p+j-1})}_{\text{buy deficit in $\text{RD}_j$}}\nonumber\\
    &+\underbrace{\sum_{k=p+j+1}^{p+q} \pi_{k} \mathbb{E}\{D_{k}\}}_{\text{buy deficit in following RDs}}+\underbrace{\pi_\ell \mathbb{E}\{D_0\}}_{\text{recharge cost}}.
\end{align}
Examining $\mathcal{J}_{p+j}$'s first order derivative with respect to $u_{p+j}$, we can conclude that
\begin{align}
    \frac{\partial \mathcal{J}_{p+j}}{\partial u_{p+j}}>0,
\end{align}
which means the greedy policy is optimal in $\text{RD}_j$.

By now we complete our proofs for all RD periods, we focus on the RU periods in the subsequent analysis.

\textbf{Threshold Policy in RU periods}\\
The induction basis is to examine the threshold policy for $\text{RU}_p$ is optimal. This is a special case of the $(M,C)$ policy in \cite{3tiercontrol}. Based on this induction basis, we can use backward induction to complete our proof. To  guarantee the optimal reservation is unique in the ramp-up periods, we need the derivative the total cost $\mathcal{J}_{j}$ in $\text{RU}_j$ and following periods with respect to $u_{j}$ is monotone increasing, therefore we make the following technical assumption.

\emph{Assumption $7$:} For each firm's demand in each non off peak period (denoted as $X$), its probability density function $f_X(x)$ is differentiable, and $f_X(x)\!>\!0$ if $x\geq0$;

\textbf{Investment decision}\\
We can regard the investment decision as a reservation decision in the virtual period $\text{RU}_0$ (off peak) with the rate $\pi_0\!=\!\pi_s+\pi_\ell$, then we get the optimal size $C^*$ which satisfies Eq. (\ref{invest}).

By now we have proved our control policy achieves maximal profit.$\hfill\blacksquare$

\subsection{Characterize N.E. of CDG}
In this subsection we characterize the N.E. of the CDG, and prove it supports maximal social welfare. Before we start, we define some useful functions for subsequent analysis.
\begin{align*}
    \delta(C_c)=\sum_{m=1}^{q} (\pi_{p+m}-\pi_{p+m+1})f_{ \{\sum_{k=1}^{p+m} X_{k}^c\}}(C_c),
\end{align*}
\begin{equation}
    \begin{aligned}
    &\gamma_i(C_c) \!=\!\sum_{j=1}^{q} \rho_j(C_c)\mathbb{E}\Bigg[ \sum_{k=1}^{p+j} X^i_k\Bigg|\sum_{k=1}^{p+j} X_k^c\!=\!C_c,\\
    &\qquad\sum_{k=1}^n X_k^c<C_c-M_n^*, 1\leq n \leq p \Bigg],
        \nonumber
    \end{aligned}
\end{equation}
and
\begin{align*}
    \rho_l(C_c)=\frac{(\pi_{p+l}-\pi_{p+l+1})f_{\{\sum_{k=1}^{p+l} X_{k}^c\}}(C_c)}{\sum_{m=1}^{q} (\pi_{p+m}-\pi_{p+m+1})f_{ \{\sum_{k=1}^{p+m} X_{k}^c\}}(C_c)}.
\end{align*}

Given other firms' investment decision $C_{-i}$, firm $i$'s optimal investment decision can be stated as:
\begin{align}
    C_i^*=\arg\min_{C_i} \mathcal{I}_i(C_i,C_{-i}).
\end{align}

$\mathcal{I}_i$ is the total electricity cost of firm $i$. 
\begin{align}
    \mathcal{I}_i(C_i,C_{-i})&=\!\!\underbrace{\pi_s C_i}_{\text{invest cost}}
    \!\!+\!\!\underbrace{\sum\nolimits_{j=1}^{p+q} \mathbb{E}\{ \pi_j^{a*} D_j^i\}}_{\text{cost in non off peak periods}}
    \!\!+\underbrace{\pi_\ell \mathbb{E}\{D_0^i\}.}_{\text{recharge cost}}
\end{align}
where $D_j^i$ is the deficit of firm $i$ in $k^{th}$ non off peak period, and $D_0^i$ is the amount of energy that firm $i$ needs to recharge its storage. Negative value means the firm shares its energy to others and obtain certain profit.

Due to our pricing mechanisms, single firm's concrete reservations at {N.E.} in each non off peak period do not influence its profit as we analyzed in Section \uppercase\expandafter{\romannumeral4}-A.
Thus, without loss of generality, we suppose firm $i$ has zero reservation in each non off peak period. Then 
\begin{align*}
    &D_1^i=x_1^i-C_i,\\
    &D_j^i=x_j^i, j=2,\cdots,p+q,\\
    &D_0^i=C_i.
\end{align*}

The derivative of $\mathcal{I}_i$ with respect to $C_i$ is
\begin{align}
    \frac{\partial \mathcal{I}_i}{\partial C_i}
    \!=\!\pi_s\!-\!\sum_{k=1}^{p+q}(\pi_k\!-\!\pi_\ell)P_0^k(C_c)\!
    -\delta(C_c)(\gamma_i(C_c)-C_i).
    \label{derioffirmi}
\end{align}
Define
\begin{align}
    \alpha&=\pi_s-\sum_{k=1}^{p+q}(\pi_k\!-\!\pi_\ell)P_0^k(C_c^*),\\
    \beta&=\delta(C_c^*).
\end{align}
%We suppose $\beta>0$ (this is a condition just for clear proof). 
Note that Assumption 7 guarantees that $\beta>0$ and $C^*>0, \forall i$. The definition of N.E. guarantees that for each firm $i$, its cost reaches minimum at $C_i^*$:
\begin{align}
    \frac{\partial \mathcal{I}_i}{\partial C_i}\bigg|_{C_i=C_i^*}=\alpha
    -\beta(\gamma_i(C_c^*)-C_i^*)=0.
    \label{equcon1}
\end{align}
Summing over all firms yields 
\begin{align}
    n\alpha-\beta(\sum\nolimits_i\gamma_i(C_c^*)-C_c^*)= 0,
\end{align}
where $n$ is the number of firms. Note that $\sum\nolimits_i \gamma_i(C_c^*)=C_c^*$, we can conclude that
\begin{align}
    \alpha= 0.
    \label{equ13}
\end{align}
Hence, Eq. (\ref{derioffirmi}) dictates that for each firm $i$,
\begin{align}
    C_i^*=\gamma_i(C_c^*).
    \label{NE}
\end{align}
This implies $C_c^*$ satisfying Eq. (\ref{Cc}), the optimal investment for the collective, in other words, the firms make investment decisions which are optimal for the collective in the N.E..$\hfill\blacksquare$
%If the N.E. exists, it is the unique $(C_1^*,c_2^*,\cdots,C_n^*)$ as we state in (\ref{NE}). Next we prove it actually exists under our technical alignment conditions. 

\subsection{Alignment Conditions for the Existence of N.E.}
We consider firm $i$'s decision $C_i$ when fixing the decisions of all
other firms $C^*_{-i}$ as
\begin{align*}
    C^*_{k}= \gamma_k(C_c^*), k\not=i.
\end{align*}

Define 
\begin{align*}
    \nu=\sum\nolimits_{k\not= i} C^*_{k}.
\end{align*}

According to (\ref{derioffirmi}), the derivative of expected daily cost of firm $i$ with respect to $C_i$ is
\begin{align}
    \frac{\partial \mathcal{I}_i}{\partial C_i}
    &=\underbrace{\pi_s-\sum_{k=1}^{p+q}(\pi_k\!-\!\pi_\ell)P_0^k(C_i+\nu)}_{\phi(C_i)}\!\nonumber\\
    &-\delta(C_i+\nu)\underbrace{(\gamma_i(C_i+\nu)-C_i)}_{\psi(C_i)}.
    \label{indpartial}
\end{align}
We obtain $\phi(C_i)$ is strictly monotone increasing with respect to $C_i$ in that
\begin{align}
    \frac{\partial^2 \phi(C_i)}{\partial C_i^2}=(\pi_{p+q}-\ell)\theta_0(C_i+\nu)>0,
\end{align}
where
\begin{align*}
    &\theta_0(C_i+\nu)=\nonumber\\
    &\int_0^{C_i+\nu-M_{1}^*} \!\!f_{X_{1}}(x_{1}) 
    \!\cdot\!\cdot\!\cdot\! \int_0^{C_i+\nu-\sum\limits_{k=1}^{p-1} x_k-M_{p}^*}f_{X_p}(x_p) \nonumber \\
    &f_{\{\sum_{k=1}^q X_{p+k}\}}(C_i\!+\!\nu\!-\!\!\sum\limits_{k=j+1}^p x_k) d x_p\!\cdot\!\cdot\!\cdot\! d x_{1}>0.
\end{align*}
Here we use Assumption 7 once more.
%and we have $\phi(C_i^*)=0$.

To verify the monotonicity of $\psi(C_i)$, we make critical use of the technical alignment conditions in (\ref{align}) to conclude
\begin{align*}
    1=\sum_k \underbrace{\frac{\partial G_{j}^k(r)}{\partial r}}_{\geq0} \Longrightarrow \frac{\partial G_{j}^i(r)}{\partial r} \leq 1.
\end{align*}
And notice that
\begin{align*}
    \sum_{j=1}^p \rho_j(C_i+\nu)=1.
\end{align*}
It then
follows that
\begin{align*}
    \frac{\partial \psi(C_i)}{\partial C_i}=\sum_{j=1}^p \rho_j(C_i+\nu)\frac{\partial G_{j}^i(C_i+\nu)}{\partial (C_i+\nu)}-1 \leq 0.
\end{align*}
This means $\psi(C_i)$ is monotonically decreasing with respect to $C_i$.
%and $\psi(C_i^*)=0$.

We obtain $\phi(C_i)$ is strictly monotone increasing with respect to $C_i$ and $\psi(C_i)$ is monotone decreasing, so $\mathcal{I}_i(C_i)$ is strictly monotone increasing according to (\ref{indpartial}). Notice $\phi(C_i^*)=\psi(C_i^*)=0$. As a result, we have
\begin{align*}
    \frac{\partial \mathcal{I}_i}{\partial C_i}\left\{
    \begin{array}{rcl}
    <0\quad C_i<C_i^*, \\
    =0\quad C_i=C_i^*, \\
    >0\quad C_i>C_i^*.
\end{array} \right.
\end{align*}
This proves that $C_i^*$ is the global minimizer of firm $i$'s cost,
establishing that $(C_1^*,c_2^*,\cdots,C_n^*)$ is the unique N.E..$\hfill\blacksquare$

\subsection{Proof of Proposition 6.2}
(a) \emph{Coalitional Stability}

For firms $\{1,2,...,n\}$, we form coalitions $\mathbb{A}_j \subset \{1,...,n\}$ such that
\begin{align*}
    \mathbb{A}_i \cap \mathbb{A}_j = \emptyset,\ \bigcup_k \mathbb{A}_k = \{1,...,n\}.
\end{align*}
The initial CDG $\mathcal{G}$ induces a new CDG $\mathcal{H}$ with players $\mathbb{A}_i$ and
associated cost
$$\mathcal{I}_{\mathbb{A}_i} = \sum_{k \in \mathbb{A}_i} \mathcal{I}_k(C_1,..,C_n).$$
Since the alignment condition (\ref{align}) holds for CDG $\mathcal{G}$, we have for any coalition $\mathbb{A}_i$,
\begin{align*}
    \frac{\partial G_{j}^{\mathbb{A}_i}(r)}{\partial r}=\sum_{k \in \mathbb{A}_i} \frac{\partial G_{j}^{k}(r)}{\partial r}\geq0.
\end{align*}
Thus, the alignment condition holds for the induced CDG $\mathcal{H}$. It therefore admits a unique N.E. $D^*$ where
\begin{align*}
    D_{\mathbb{A}_i}^* =\gamma_{\mathbb{A}_i}(C_c^*)=\sum_{k \in \mathbb{A}_i} \gamma_k(C_c^*).
\end{align*} 
Now individual rationality of $D^*$ in CDG $\mathcal{H}$ is equivalent
to coalitional stability of CDG $\mathcal{G}$, proving the claim.

(b) If a firm $i$ has no demand, then $C_i^*=0$ according to (\ref{indinvest}).
$\hfill\blacksquare$
}
\end{document}